\documentstyle[12pt]{article}
\begin{document}

\title{{{\bf Reconstructing $f(T)$-gravity from the polytropic and different Chaplygin gas dark energy models}}}

%\title{{{\bf Original and entropy-corrected versions of the holographic and new agegraphic $f(T)$-gravity models}}}

%\title{{{\bf Reconstructing $f(T)$-gravity from the holographic and new agegraphic dark energy}}}

\author{K. Karami$^{1,2}$\thanks{E-mail: KKarami@uok.ac.ir} , A. Abdolmaleki$^{1}$\thanks{E-mail: AAbdolmaleki@uok.ac.ir}\\
$^{1}$\small{Department of Physics, University of Kurdistan,
Pasdaran St., Sanandaj, Iran}\\$^{2}$\small{Research Institute for
Astronomy
$\&$ Astrophysics of Maragha (RIAAM), Maragha, Iran}\\
}

\maketitle

\begin{abstract}
Motivated by a recent work of us \cite{Karami1}, we reconstruct the
different $f(T)$-gravity models corresponding to a set of dark
energy scenarios containing the polytropic, the standard Chaplygin,
the generalized Chaplygin and the modified Chaplygin gas models. We
also derive the equation of state parameter of the selected
$f(T)$-gravity models and obtain the necessary conditions for
crossing the phantom-divide line.
\end{abstract}


\newpage

%--------------------------------------------------------------------------------------------------------------------
\section{Friedmann equations in the $f(T)$ theory}

In the framework of $f(T)$-gravity, the Friedmann equations in the
flat spatial Friedmann-Robertson-Walker (FRW) universe are given by
\cite{Karami1,Ferraro,bengochea,yerzhanov,linder,WufT,Bamba}

\begin{equation}
\frac{3}{k^2}H^2=\rho+\rho_T,\label{fT11}
\end{equation}
\begin{equation}
\frac{1}{k^2}(2\dot{H}+3H^2)=-(p+p_T),\label{fT22}
\end{equation}
where
\begin{equation}
\rho_T=\frac{1}{2k^2}(2Tf_T-f-T),\label{roT}
\end{equation}
\begin{equation}
p_T=-\frac{1}{2k^2}[-8\dot{H}Tf_{TT}+(2T-4\dot{H})f_T-f+4\dot{H}-T],\label{pT}
\end{equation}
with
\begin{equation}
T=-6H^2.\label{T}
\end{equation}
Here $k^2=8\pi G$, $H=\dot{a}/a$ is the Hubble parameter, $\rho$ and
$p$ are the total energy density and pressure of the matter inside
the universe. Also $\rho_T$ and $p_T$ are the energy density and
pressure due to the contribution of the torsion scalar $T$.
Furthermore, in $f_T$ and $f_{TT}$ the subscript $T$ denotes a
derivative with respect to $T$. Note that if $f(T)=T$ then Eqs.
(\ref{fT11}) and (\ref{fT22}) transform to the usual Friedmann
equations in Einstein general relativity (GR).

The energy conservation laws are
\begin{equation}
\dot{\rho}+3H(\rho+p)=0,
\end{equation}
\begin{equation}
\dot{\rho}_T+3H(\rho_T+p_T)=0.\label{ecT}
\end{equation}
The equation of state (EoS) parameter due to the torsion
contribution is defined as \cite{Karami1,yerzhanov}
\begin{equation}
\omega_T=\frac{p_T}{\rho_T}=-1+\frac{8\dot{H}Tf_{TT}+4\dot{H}f_T-4\dot{H}}{2Tf_T-f-T}.\label{omegaT}
\end{equation}

For a given $a=a(t)$, by the help of Eqs. (\ref{roT}) and (\ref{pT})
one can reconstruct the $f(T)$-gravity according to any dark energy
(DE) model given by the EoS $p_T=p_T(\rho_T)$ or $\rho_T=\rho_T(a)$.

Here we assume a pole-like phantom scale factor as \cite{Sadjadi}
\begin{equation}
a(t)=a_0(t_s-t)^{-h},~~~t\leq t_s,~~~h>0.\label{a}
\end{equation}
Using Eqs. (\ref{T}) and (\ref{a}) one can obtain
\begin{equation}
\begin{array}{l}
 H=\frac{h}{t_s-t},\\
T=-\frac{6h^2}{(t_s-t)^2},\\
 \dot{H}=-\frac{T}{6h}.\label{respect to r}
\end{array}
\end{equation}
From Eqs. (\ref{a}) and (\ref{respect to r}) the scale factor $a$
can be rewritten in terms of $T$ as
\begin{equation}
a=a_0{\left(-\frac{T}{6h^2}\right)}^{\frac{h}{2}}.\label{aT}
\end{equation}

%--------------------------------------------------------------------------------------------------------------------
\section{Polytropic $f(T)$-gravity model}

Here like \cite{Karami1} we reconstruct the $f(T)$-gravity from the
polytropic gas DE model. Following \cite{Karami2}, the EoS of the
polytropic gas is given by
\begin{equation}
p_{\Lambda}=K\rho_{\Lambda}^{1+\frac{1}{n}},\label{pol1}
\end{equation}
where $K$ is a positive constant and $n$ is the polytropic index.
Using Eq. (\ref{ecT}) the energy density evolves as
\begin{equation}
\rho_{\Lambda}=\left(Ba^\frac{3}{n}-K\right)^{-n},\label{pol2}
\end{equation}
where $B$ is a positive integration constant \cite{Karami2}.

Replacing Eq. (\ref{aT}) into (\ref{pol2}) yields
\begin{equation}
\rho_{\Lambda}=\left(\alpha
T^\frac{3h}{2n}-K\right)^{-n},\label{rhoPG}
\end{equation}
where
\begin{equation}
\alpha=Ba_{0}^{\frac{3}{n}}{\left(-6h^2\right)}^{\frac{-3h}{2n}}.\label{alphaP}
\end{equation}
Equating (\ref{roT}) with (\ref{rhoPG}), i.e.
$\rho_T=\rho_{\Lambda}$, we obtain the following differential
equation
\begin{equation}
2Tf_T-f-T-2k^2\left(\alpha
T^\frac{3h}{2n}-K\right)^{-n}=0.\label{dif eq1}
\end{equation}
Solving Eq. (\ref{dif eq1}) gives
\begin{equation}
f(T)=\beta~T^{1/2}+T+(-1)^{1+n}\frac{2k^2}{K^n}
~{_2}F_1\left(-\frac{n}{3h},n;1-\frac{n}{3h};\frac{\alpha}{K}T^\frac{3h}{2n}\right),\label{fPDE}
\end{equation}
where $_{2}F_1$ denotes the first hypergeometric function. Replacing
Eq. (\ref{fPDE}) into (\ref{omegaT}) one can
 obtain the EoS parameter of torsion contribution as
\begin{equation}
\omega_{T}=-1-\frac{1}{\frac{K}{\alpha}T^{\frac{-3h}{2n}}-1}~,~~~h>0.\label{wHDE}
\end{equation}
Using Eqs. (\ref{T}) and (\ref{alphaP}), the above relation can be
rewritten as
\begin{equation}
\omega_{T}=-1-\frac{1}{\frac{K}{B}\left[a_0\left(\frac{H}{h}\right)^{h}\right]^{\frac{-3}{n}}-1}~,~~~h>0.\label{wCGDE}
\end{equation}
We see that for
$\frac{K}{B}\left[a_0\left(\frac{H}{h}\right)^{h}\right]^{\frac{-3}{n}}>1$,
$\omega_T<-1$ which corresponds to a phantom accelerating
universe.
%--------------------------------------------------------------------------------------------------------------------
\section{Standard Chaplygin $f(T)$-gravity model}

The EoS of the standard Chaplygin gas (SCG) DE is given by
\cite{Kamenshchik}
\begin{equation}
p_{\Lambda}=-\frac{A}{\rho_{\Lambda}},
\end{equation}
where $A$ is a positive constant. Inserting the above EoS into the
energy conservation equation (\ref{ecT}), leads to a density
evolving as \cite{Kamenshchik}
\begin{equation}
\rho_{\Lambda}=\sqrt{A+\frac{B}{a^6}},\label{CG}
\end{equation}
where $B$ is an integration constant.

Inserting Eq. (\ref{aT}) into (\ref{CG}) one can get
\begin{equation}
\rho_{\Lambda}=\sqrt{A+\alpha T^{-3h}},\label{rhoCG}
\end{equation}
where
\begin{equation}
\alpha=B{a_0}^{-6}{(-6h^2)}^{3h}.\label{alphaCG}
\end{equation}
Equating (\ref{rhoCG}) with (\ref{roT}) one can obtain
\begin{equation}
2Tf_T-f-T-2k^2\sqrt{A+\alpha T^{-3h}}=0.\label{dif eq2}
\end{equation}
Solving the differential equation (\ref{dif eq2}) yields
\begin{eqnarray}
f(T)=\beta T^{1/2}+T-2k^2A^{\frac{1}{2}}~
{_2}F_1\left(\frac{1}{6h},\frac{-1}{2};1+\frac{1}{6h};{-\frac{\alpha}{A}}
T^{-3h}\right).\label{fCGDE}
\end{eqnarray}
Replacing Eq. (\ref{fCGDE}) into (\ref{omegaT}) one can
 get
\begin{equation}
\omega_{T}=-1+\frac{1}{\frac{A}{\alpha}T^{3h}+1}~,~~~h>0.\label{wCGDE}
\end{equation}
Using Eqs. (\ref{T}) and (\ref{alphaCG}), the above relation can
be rewritten as
\begin{equation}
\omega_{T}=-1+\frac{1}{\frac{A}{B}\left[a_0\left(\frac{H}{h}\right)^{h}\right]^{6}+1}~,~~~h>0,\label{wCGDE}
\end{equation}
which for $B<0$ and
$\frac{A}{|B|}\left[a_0\left(\frac{H}{h}\right)^{h}\right]^{6}>1$
then $\omega_T$ can cross the phantom-divide line.
%--------------------------------------------------------------------------------------------------------------------
\section{Generalized Chaplygin $f(T)$-gravity model}

The EoS of the Generalized Chaplygin Gas (GCG) DE model is given
by \cite{Bento}
\begin{equation}
p_{\Lambda}=-\frac{A}{\rho_{\Lambda}^\alpha},
\end{equation}
where $\alpha$ is a constant in the range $0\leq\alpha\leq 1$ (the
SCG corresponds to the case $\alpha=1$) and $A$ a positive constant.
Using Eq. (\ref{ecT}), the GCG energy density evolves as
\cite{Bento}
\begin{equation}
\rho_{\Lambda}=\left({A+\frac{B}{a^{3(1+\alpha)}}}\right)^{\frac{1}{1+\alpha}},\label{GCG}
\end{equation}
where $B$ is an integration constant.

Substituting Eq. (\ref{aT}) into (\ref{GCG}) one can get
\begin{equation}
\rho_{\Lambda}={\left(A+\gamma~
T^{\frac{-3}{2}h(1+\alpha)}\right)}^{\frac{1}{1+\alpha}},\label{rhoGCG}
\end{equation}
where
\begin{equation}
\gamma=B{a_0}^{-3(1+\alpha)}{\left(-6h^2\right)}^{\frac{3}{2}h(1+\alpha)}.\label{gammaGCG}
\end{equation}
Equating (\ref{rhoGCG}) with (\ref{roT}) gives
\begin{equation}
2Tf_T-f-T-2k^2{\left({A+\gamma~
T^{\frac{-3}{2}h(1+\alpha)}}\right)}^{\frac{1}{1+\alpha}}=0.\label{dif
eq3}
\end{equation}
Solving Eq. (\ref{dif eq3}) yields
\begin{eqnarray}
f(T)=\beta T^{1/2}+T-2k^2A^{\frac{1}{1+\alpha}}~
{_2}F_1\left(\frac{1}{3h(1+\alpha)},\frac{-1}{1+\alpha};1+\frac{1}{3h(1+\alpha)};{-\frac{\gamma}{A}}
T^{\frac{-3}{2}h(1+\alpha)}\right),\label{fGCG}
\end{eqnarray}
Replacing Eq. (\ref{fGCG}) into (\ref{omegaT}) gives the EoS
parameter as
\begin{equation}
\omega_{T}=-1+\frac{1}{\frac{A}{\gamma}T^{\frac{3}{2}h(1+\alpha)}+1}~,~~~h>0,~~~0\leq\alpha\leq1,\label{wCGDE}
\end{equation}
Using Eqs. (\ref{T}) and (\ref{gammaGCG}), the above relation can
be rewritten as
\begin{equation}
\omega_{T}=-1+\frac{1}{\frac{A}{B}\left[a_0\left(\frac{H}{h}\right)^{h}\right]^{3(1+\alpha)}+1},~~~h>0,~~~0\leq\alpha\leq1,\label{wCGDE}
\end{equation}
which for $B<0$ and
$\frac{A}{|B|}\left[a_0\left(\frac{H}{h}\right)^{h}\right]^{3(1+\alpha)}>1$
then $\omega_T$ can cross the phantom-divide line.
%--------------------------------------------------------------------------------------------------------------------
\section{Modified Chaplygin $f(T)$-gravity model}

The EoS of the modified Chaplygin gas (MCG) DE model is given by
\cite{Benaoum}
\begin{equation}
p_{\Lambda}=A\rho_\Lambda-\frac{B}{\rho_{\Lambda}^\alpha},
\end{equation}
where $A$ and $B$ are positive constants and $0\leq\alpha\leq1$.
Using Eq. (\ref{ecT}), the MCG energy density evolves as
\cite{Benaoum}
\begin{equation}
\rho_{\Lambda}=\left({\frac{B}{1+A}+\frac{C}{a^{3(1+\alpha)(1+A)}}}\right)^{\frac{1}{1+\alpha}},\label{MCG}
\end{equation}
where $C$ is an integration constant.

Replacing Eq. (\ref{aT}) into (\ref{MCG}) yields
\begin{equation}
\rho_{\Lambda}=\left({\frac{B}{1+A}+\gamma~
T^{\frac{-3}{2}h(1+\alpha)(1+A)}}\right)^{\frac{1}{1+\alpha}},\label{rhoMCG}
\end{equation}
where
\begin{equation}
\gamma=C{a_0}^{-3(1+\alpha)(1+A)}{\left(-6h^2\right)}^{\frac{3}{2}h(1+\alpha)(1+A)}.\label{gammaMCG}
\end{equation}
Equating (\ref{rhoGCG}) with (\ref{roT}) gives
\begin{equation}
2Tf_T-f-T-2k^2\left({\frac{B}{1+A}+\gamma~
T^{\frac{-3}{2}h(1+\alpha)(1+A)}}\right)^{\frac{1}{1+\alpha}}=0.\label{dif
eq4}
\end{equation}
Solving Eq. (\ref{dif eq4}) yields
\begin{eqnarray}
f(T)=\beta~T^{1/2}+T-2k^2\left(\frac{B}{1+A}\right)^{\frac{1}{1+\alpha}}
~~~~~~~~~~~~~~~~~~~~~~~~~~~~~~~~~~~~~~~~~~~~~~~~~~~~~~~~~~~~~~~
\nonumber\\\times{_2}F_1\left(\frac{1}{3h(1+\alpha)(1+A)},\frac{-1}{1+\alpha};1+\frac{1}{3h(1+\alpha)(1+A)};{\frac{-\gamma(1+A)}{B}}
T^{\frac{-3}{2}h(1+\alpha)(1+A)}\right).\label{fMCG}
\end{eqnarray}
 Replacing Eq. (\ref{fMCG}) into (\ref{omegaT}) one can
 obtain the EoS parameter of torsion contribution as
\begin{equation}
\omega_{T}=-1+\frac{A+1}{\frac{B}{\gamma(1+A)}T^{\frac{3}{2}h(1+\alpha)(1+A)}+1}~,~~~h>0,~~~0\leq\alpha\leq1,\label{wCGDE}
\end{equation}
and using Eqs. (\ref{T}) and (\ref{gammaMCG}), it can be rewritten
as
\begin{equation}
\omega_{T}=-1+\frac{A+1}{\frac{B}{C(1+A)}\left[a_0\left(\frac{H}{h}\right)^{h}\right]^{3(1+\alpha)(1+A)}+1},~~~h>0,~~~0\leq\alpha\leq1,\label{wCGDE}
\end{equation}
which for $C<0$ and
$\frac{B}{|C|(1+A)}\left[a_0\left(\frac{H}{h}\right)^{h}\right]^{3(1+\alpha)(1+A)}>1$
then $\omega_T$ can cross the phantom-divide line.
%--------------------------------------------------------------------------------------------------------------------
\section{Conclusions}
Here we considered the polytropic gas, the SCG, the GCG and the MCG
models of the DE. We reconstructed the different theories of
modified gravity based on the $f(T)$ action in the spatially-flat
FRW universe and according to the selected DE models. We also
obtained the EoS parameter of the polytropic, standard Chaplygin,
generalized Chaplygin and modified Chaplygin $f(T)$-gravity
scenarios. We showed that crossing the phantom-divide line can occur
when the constant parameters of the models to be chosen properly.
%--------------------------------------------------------------------------------------------------------------------
\\
\\
\noindent{{\bf Acknowledgements}}\\
The work of K. Karami has been supported financially by Research
Institute for Astronomy $\&$ Astrophysics of Maragha (RIAAM),
Maragha, Iran.
%--------------------------------------------------------------------------------------------------------------------

%--------------------------------------------------------------------------------------------------------------------
\end{document}